\documentclass[twocolumn,amsmath,amssymb,floatfix,nature,shownopacs,footinbib,superscriptaddress]{revtex4-1}
\usepackage{amsmath,amssymb,natbib,bm,graphicx,url,epsf,color}
\usepackage[british]{babel}
\usepackage{wasysym}
\usepackage{MnSymbol}
\setcitestyle{super}

\begin{document}

\title{Heat Coulomb Blockade of One Ballistic Channel}

\author{E. Sivre}
\affiliation{Centre de Nanosciences et de Nanotechnologies (C2N), CNRS, Univ Paris Sud-Universit\'e Paris-Saclay, 91120 Palaiseau, France}
\author{A. Anthore}
\affiliation{Centre de Nanosciences et de Nanotechnologies (C2N), CNRS, Univ Paris Sud-Universit\'e Paris-Saclay, 91120 Palaiseau, France}
\affiliation{Univ Paris Diderot-Sorbonne Paris Cit\'e}
\author{F.D. Parmentier}
\affiliation{Centre de Nanosciences et de Nanotechnologies (C2N), CNRS, Univ Paris Sud-Universit\'e Paris-Saclay, 91120 Palaiseau, France}
\author{A. Cavanna}
\affiliation{Centre de Nanosciences et de Nanotechnologies (C2N), CNRS, Univ Paris Sud-Universit\'e Paris-Saclay, 91120 Palaiseau, France}
\author{U. Gennser}
\affiliation{Centre de Nanosciences et de Nanotechnologies (C2N), CNRS, Univ Paris Sud-Universit\'e Paris-Saclay, 91120 Palaiseau, France}
\author{A. Ouerghi}
\affiliation{Centre de Nanosciences et de Nanotechnologies (C2N), CNRS, Univ Paris Sud-Universit\'e Paris-Saclay, 91120 Palaiseau, France}
\author{Y. Jin}
\affiliation{Centre de Nanosciences et de Nanotechnologies (C2N), CNRS, Univ Paris Sud-Universit\'e Paris-Saclay, 91120 Palaiseau, France}
\author{F. Pierre\thanks{frederic.pierre@u-psud.fr}}
\email[e-mail: ]{frederic.pierre@u-psud.fr}
\affiliation{Centre de Nanosciences et de Nanotechnologies (C2N), CNRS, Univ Paris Sud-Universit\'e Paris-Saclay, 91120 Palaiseau, France}

\maketitle

{\sffamily
Quantum mechanics and Coulomb interaction dictate the behavior of small circuits.
The thermal implications cover fundamental topics from quantum control of heat to quantum thermodynamics, with prospects of novel thermal machines and an ineluctably growing influence on nanocircuit engineering\cite{Pekola2015,Vinjanampathy2016}.
Experimentally, the rare observations thus far include the universal thermal conductance quantum\cite{Schwab2000,Meschke2006,Jezouin2013b,Banerjee2017,Cui2017} and heat interferometry\cite{Giazotto2012}.
However, evidences for many-body thermal effects paving the way to markedly different heat and electrical behaviors in quantum circuits remain wanting.
Here we report on the observation of the Coulomb blockade of electronic heat flow from a small metallic circuit node, beyond the widespread Wiedemann-Franz law paradigm.
We demonstrate this thermal many-body phenomenon for perfect (ballistic) conduction channels to the node, where it amounts to the universal suppression of precisely one quantum of conductance for the transport of heat, but none for electricity\cite{Slobodeniuk2013}.
The inter-channel correlations that give rise to such selective heat current reduction emerge from local charge conservation, in the floating node over the full thermal frequency range ($\lesssim$temperature$\times k_\mathrm{B}/h$).
This observation establishes the different nature of the quantum laws for thermal transport in nanocircuits.
}

\begin{figure}[!tbh]
\renewcommand{\figurename}{\textbf{Figure}}
\renewcommand{\thefigure}{\textbf{\arabic{figure}}}
\centering\includegraphics[width=89mm]{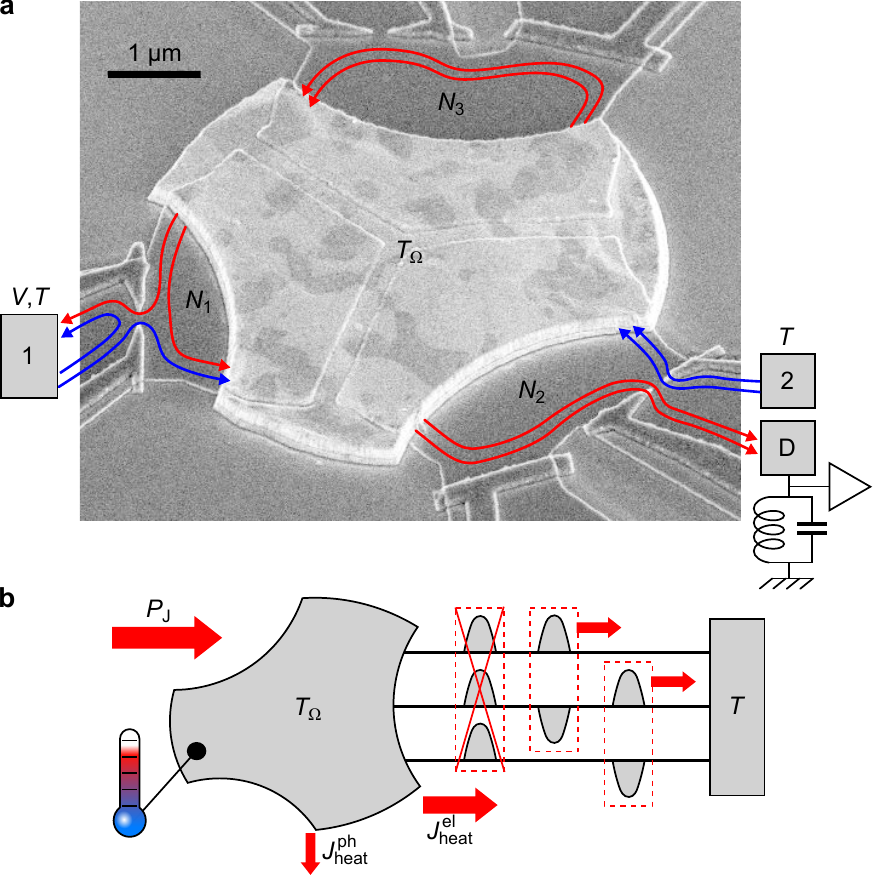}
\caption{
\footnotesize
\textbf{Experimental setup.}
\textbf{a}, Device micrograph.
A small metallic island (brighter) is in galvanic contact with three distinct branches of a two-dimensional electron gas (darker grey; etched trenches visible underneath the island).
The connection to large electrodes further away (represented by rectangles for branches 1 and 2) is controlled by field effect using capacitively coupled gates (grey with a bright delimitation).
The sample is set in the integer quantum Hall regime at filling factor $\nu=2$, where the current propagates along two chiral edge channels (lines with arrows).
In the displayed configuration, electrodes 1, 2 and 3 are connected by, respectively, one, two and zero fully transmitted channels ($N_1=1, N_2=2$, $N_3=0$).
Applying a dc bias voltage $V$ to electrode 1 dissipates the Joule power $P_\mathrm{J}$ into the island.
The resulting temperature rise $T_\Omega-T$ is determined from the measured increase of electrical fluctuations on electrode D.
\textbf{b}, Heat flow schematic.
Injected power and net outgoing heat current exactly compensate each other in the steady state ($P_\mathrm{J}=J_\mathrm{heat}^\mathrm{el}+J_\mathrm{heat}^\mathrm{ph}$).
The $N$ ballistic electronic channels (here $N=3$ shown as black lines) can be mapped onto one channel-symmetric charge mode suppressed by the heat Coulomb blockade (crossed symmetric charge pulses), and $N-1$ independent dipole (neutral) modes decoupled from the island's charge (antisymmetric charge pulses).
\normalsize
}
\label{fig1}
\end{figure}

The non-interacting `scattering' approach to quantum transport describes coherent conductors as a set of independent channels\cite{Landauer1975,Buttiker1986}.
However, in circuits with small floating nodes, the Coulomb interaction induces inter-channel correlations, including among distinct conductors connected to the same node.
Consequences are wide-ranging, from the emblematic `Coulomb blockade' suppression of electrical conduction at low voltages and temperatures\cite{Kulik1975,Averin1986,Nazarov1989,Ingold1992} to exotic `charge' Kondo physics\cite{Matveev1991,Iftikhar2015}.
Remarkably, Coulomb effects can be profoundly different in the charge and heat sectors, in violation of the standard Wiedemann-Franz ratio between electronic conductances of heat and electricity ($\pi^2k_\mathrm{B}^2T/3e^2$ with $e$ the elementary electron charge, $k_\mathrm{B}$ the Boltzmann constant, $T$ the temperature).
For ballistic conductors, along which electrons are never reflected backward, the electrical conductance $G_\mathrm{elec}$ is predicted\cite{Flensberg1993,Yeyati2001,Kindermann2003} and found\cite{Altimiras2007,Parmentier2011,Jezouin2013} immune against Coulomb blockade, essentially because charge flow is noiseless ($G_\mathrm{elec}=N\times G_\mathrm{Q}^\mathrm{e}$, with $N$ the number of channels, $G_\mathrm{Q}^\mathrm{e}=e^2/h$ the electrical conductance quantum and $h$ the Planck constant).
Nonetheless, theory predicts\cite{Slobodeniuk2013} a universal suppression of the heat conductance $G_\mathrm{heat}$ across ballistic conductors connected to a small, floating circuit node by precisely one quantum of thermal conductance $G_\mathrm{Q}^\mathrm{h}=\pi^2k_\mathrm{B}^2T/3h$ ($G_\mathrm{heat}=(N-1)\times G_\mathrm{Q}^\mathrm{h}$), as presently observed experimentally.

This violation of the Wiedemann-Franz relation does not result from an energy dependent electronic density of states, nor from the high-pass energy filtering across single electron transistors\cite{Altimiras2012,Dutta2017}.
We describe the underlying mechanism in the spirit of Ref.~\citenum{Slobodeniuk2013}, specifically focusing on a metallic node connected to large voltage biased electrodes through a total of $N$ ballistic channels (Fig.~1a,b).
Electronic heat currents can be viewed as the propagation of electrical current fluctuations within a broad frequency bandwidth, extending up to the thermal cutoff $\sim k_\mathrm{B}T/h$.
For a voltage biased electrode, the emitted fluctuations result from the thermal broadening of the electron Fermi distribution.
For a floating circuit node, charge conservation imposes that the thermal emission of a net charge through a current pulse is also accompanied by an opposite charge accumulation in the node.
Such charge accumulation relaxes in the characteristic $RC$ time (with $R=1/NG_\mathrm{Q}^\mathrm{e}$ here, and $C$ the node geometrical capacitance), which suppresses the overall (net) charge fluctuations emitted from the node (thermal plus subsequent relaxation) at frequencies below $\sim1/RC$. 
At low temperature $k_\mathrm{B}T\ll h/RC$ ($k_\mathrm{B}T\ll NE_\mathrm{C}$ here, with $E_\mathrm{C}\equiv e^2/2C$ the node charging energy), where this suppression covers the full thermal frequency range, it can result in an important reduction of the total heat current.
Note that the electrical fluctuations emitted from such a floating node were previously explored in the context of using a `voltage probe' to emulate inelastic mechanisms within the scattering theory of quantum transport (see e.g. Ref.~\citenum{Blanter2000} and references within).
An intuitive way to understand why, for ballistic conductors this reduction amounts universally to one heat transport channel, is to note\cite{Slobodeniuk2013} that $N$ ballistic electronic channels can be mapped onto a single charge mode (e.g. identical current fluctuations on all electronic channels) and $N-1$ independent neutral modes (e.g. opposite current fluctuations on each of $N-1$ pairs of electronic channels), as schematically illustrated Fig.~1b.
The $N-1$ ballistic neutral modes are completely decoupled from the global charge of the island, and therefore contribute each by one (universal) quantum of thermal conductance $G_\mathrm{Q}^\mathrm{h}$.
In contrast, the electrical current fluctuations propagating along the single charge mode are directly connected with fluctuations of the node's charge.
Negligible charge accumulation in the node for frequencies $\lesssim k_\mathrm{B}T/h$ therefore completely blocks the charge mode and suppresses its contribution to heat transport, resulting in $G_\mathrm{heat}=(N-1)\times G_\mathrm{Q}^\mathrm{h}$.

The experiment expands on an approach introduced to measure the thermal conductance quantum across electronic channels\cite{Jezouin2013b}.
In contrast to previous works\cite{Jezouin2013b,Banerjee2017}, the present implementation down to electronic temperatures of 8\,mK (Methods) allows for the direct observation of the heat Coulomb blockade, which requires that energy transfers between electrons and phonons in the node remain negligible with respect to those through one ballistic channel.
The device consists of a central metallic island (the circuit node, see Fig.~1a) separately connected to three large electrodes indexed by $i\in\{1,2,3\}$ through, respectively, $N_i$ ballistic quantum channels ($N=N_1+N_2+N_3$).
A Joule power $P_\mathrm{J}=V^2G_\mathrm{P}/2$ controlled by the dc voltage $V$ applied to electrode 1 dissipates into the electronic fluid within the island (the remaining voltage generator power going into the large electrodes, see Methods), with electrodes 2 and 3 being grounded and $G_\mathrm{P}$ the corresponding conductance across the device (here, $G_\mathrm{P}=G_\mathrm{Q}^\mathrm{e}/\left[N_1^{-1}+(N_2+N_3)^{-1}\right]$).
As a result, the metallic island heats up to a steady-state electronic temperature $T_\Omega>T$ (with $T$ the base electronic temperature), such that $P_\mathrm{J}$ and net outgoing heat flow $J_\mathrm{heat}$ exactly compensate ($P_\mathrm{J}=J_\mathrm{heat}$).
The determination of $T_\Omega$ through thermal noise measurements therefore directly provides the $J_\mathrm{heat}$-$(T_\Omega-T)$ characteristics.
In general, $J_\mathrm{heat}$ is the sum of different contributions, including mainly the electronic heat current $J_\mathrm{heat}^\mathrm{el}(N,T_\Omega,T)$ and thermal transfers to the substrate phonons $J_\mathrm{heat}^\mathrm{ph}(T_\Omega,T)$:
$J_\mathrm{heat}\simeq J_\mathrm{heat}^\mathrm{el}+J_\mathrm{heat}^\mathrm{ph}$.
However, at $T_\Omega<20$\,mK, the rapidly decreasing energy transfers toward phonons are found to become negligible such that $P_\mathrm{J}=J_\mathrm{heat}\simeq J_\mathrm{heat}^\mathrm{el}$.

\begin{figure*}[!tbh]
\renewcommand{\figurename}{\textbf{Figure}}
\renewcommand{\thefigure}{\textbf{\arabic{figure}}}
\centering\includegraphics[width=136mm]{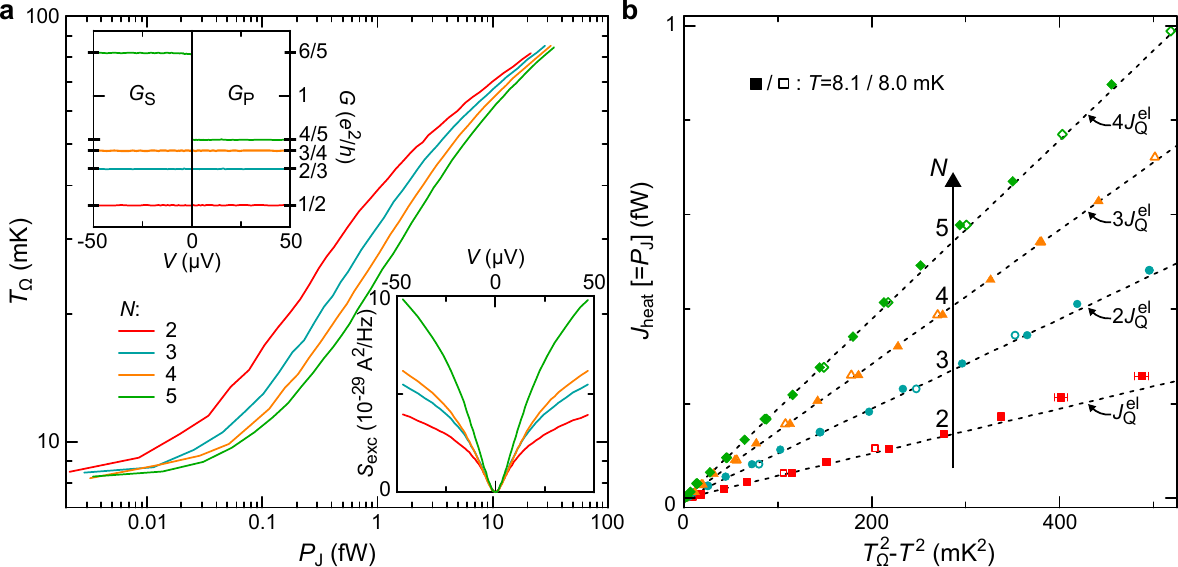}
\caption{
\footnotesize
\textbf{Heat Coulomb blockade of one ballistic channel.}
\textbf{a}, The island electron temperature $T_\Omega$ is plotted versus dissipated Joule power $P_\mathrm{J}$, for different numbers $N$ of connected ballistic channels.
It is obtained from the excess noise spectral density $S_\mathrm{exc}$ measured electrode D, shown in the bottom-right inset versus the dc voltage $V$ applied to electrode 1: $T_\Omega=T+S_\mathrm{exc}/(2k_\mathrm{B}G_\mathrm{S})$, $P_\mathrm{J}=V^2G_\mathrm{P}/2$.
Top-left inset, the device electrical conductances $G_\mathrm{S,P}$ (lines, see text) match their expected quantum limited values (superimposed thick ticks on side axis) independently of $V$: they are not reduced by Coulomb blockade.
\textbf{b}, Symbols (statistical uncertainties shown when distinctly larger) represent the overall heat flow ($J_\mathrm{heat}=P_\mathrm{J}$) displayed versus $T_\Omega^2-T^2$, at low temperatures where electron-phonon interactions are reduced ($T_\Omega<25$\,mK).
The nearby straight dashed lines show $(N-1)\times J_\mathrm{Q}^\mathrm{el}$, corresponding to a systematic heat current suppression of $1\times J_\mathrm{Q}^\mathrm{el}$.
\normalsize
}
\label{fig2}
\end{figure*}

The ballistic electronic channels are realized in a high-mobility Al(Ga)As two-dimensional electron gas (2DEG), tuned in the integer quantum Hall regime to the filling factor $\nu=2$ with an applied perpendicular magnetic field $B\simeq4.1$\,T.
Without loss of generality, we benefit from the topologically protected ballistic character of the chiral quantum Hall channels (lines with arrows in Fig.~1a).
With $\nu$ channels propagating along each edge, it is possible to adjust separately $N_i\in\{0,1,2\}$ using the field effect.
The device was tuned to $N\in\{2,3,4,5\}$ with at least one ballistic channel connecting the central island to both the large electrodes one and two ($N_1=1$, $N_2\in\{1,2\}$ and $N_3\in\{0,1,2\}$, except for tests in Methods).
The micrometer-scale metallic island, mainly composed of gold, is associated with a continuous electronic density of states and a charging energy $E_\mathrm{C}\simeq k_\mathrm{B}\times0.3$\,K (Methods).
By thermal annealing, we achieve negligible levels of electron reflection probability at the 2DEG/island interface, as detailed per channel in Methods.
This is essential not only to reach the ballistic limit, but also because residual reflections would impede the noise thermometry through additional quantum shot noise.
The electron temperature is obtained from electrical current fluctuations, resolved down to a statistical uncertainty of $\pm5\,10^{-32}$\,A$^2/$Hz at frequencies near 1\,MHz, using a homemade preamplifier\cite{Liang2012} cooled to $3.9$\,K.
Quantum shot noise measurements provide the in-situ calibrations of the noise amplification chain gain ($\pm0.1\%$, all uncertainties and displayed error bars are statistical standard errors), and of the base electronic temperature $T\simeq8$\,mK ($\pm1\%$, further confirmed by dynamical Coulomb blockade thermometry\cite{Iftikhar2016}, see Methods).
At non-zero Joule power, the temperature increase $T_\Omega-T$ of the metallic island electrons is directly connected to a rise in emitted current noise.
In the spirit of the robust Johnson-Nyquist thermometry, the excess noise $S_\mathrm{exc}$ measured on the floating electrode D (Fig.~1a) with respect to the noise at $V=0$ reads\cite{Blanter2000chaotic,Blanter2000,Jezouin2013b}:
$S_\mathrm{exc}=2G_\mathrm{S}k_\mathrm{B}(T_\Omega-T)$, with $G_\mathrm{S}$ the electrical conductance across the device from electrode 2 to electrodes 1 and 3 (in absence of electrical Coulomb blockade, $G_\mathrm{S}=G_\mathrm{Q}^\mathrm{e}/\left[N_2^{-1}+(N_1+N_3)^{-1}\right]$).
Note that further tests made in order to eliminate possible experimental artifacts are described in Methods.

In Fig.~2a we show measurements of the excess noise spectral density $S_\mathrm{exc}$ (bottom-right inset) and of the conductances $G_\mathrm{S}$ (left side of top-left inset) and $G_\mathrm{P}$ (right side of top-left inset), all as a function of the dc voltage $V$ applied to electrode~1.
Each color corresponds to one device configuration $N\,[N_1,N_2,N_3]\in\{2\,[1,1,0],$ $3\,[1,1,1],$ $4\,[1,1,2],$ $5\,[1,2,2]\}$, with a color code fixed from now on.
The island heating is manifested in the increase of $S_\mathrm{exc}$ at finite $V$.
In contrast, both $G_\mathrm{S}$ and $G_\mathrm{P}$ remain indistinguishable from their maximum quantum limit (respectively $G_\mathrm{Q}^\mathrm{e}/\left[N_2^{-1}+(N_1+N_3)^{-1}\right]$ and $G_\mathrm{Q}^\mathrm{e}/\left[N_1^{-1}+(N_2+N_3)^{-1}\right]$, displayed as thick ticks), independently of $V$.
This demonstrates the absence of Coulomb blockade reduction of the electrical conductance across ballistic channels, at an experimental accuracy better than $0.1\%$ (see also Refs~\citenum{Altimiras2007,Parmentier2011,Jezouin2013}).
Lines in the main panel of Fig.~2a represent, in a $\log$-$\log$ scale, the electron temperature in the metallic island, $T_\Omega$, versus the injected Joule power $P_\mathrm{J}$, with $T_\Omega$ obtained from $S_\mathrm{exc}$ and the separately calibrated base temperature $T\simeq8.1$\,mK. 
As generally expected, $T_\Omega$ is higher when there are fewer electronic channels to evacuate the dissipated $P_\mathrm{J}$.
Figure~2b displays as symbols the same data, as well as a subsequent run at $T\simeq8.0$\,mK, but now as the net heat flow $J_\mathrm{heat}=P_\mathrm{J}$ versus $T_\Omega^2-T^2$, and focusing on low temperatures ($T_\Omega<25$\,mK) where the phonon contribution $J_\mathrm{heat}^\mathrm{ph}$ is reduced and where a full suppression of one electronic thermal channel is predicted\cite{Slobodeniuk2013} (Methods).
The straight dashed line closest to the data for $N$ ballistic channels is $(N-1)\times J_\mathrm{Q}^\mathrm{el}$, with $J_\mathrm{Q}^\mathrm{el}=\pi^2k_\mathrm{B}^2(T_\Omega^2-T^2)/6h$ the quantum limit of heat flow per electronic channel.
The mere observation that $J_\mathrm{heat}$ is well below predictions for $N$ independent ballistic channels ($N\times J_\mathrm{Q}^\mathrm{el}$) directly demonstrates the specific suppression of heat transport from small circuit nodes, whereas electrical transport remains at the maximum quantum limit. 
Moreover, we find at such low-temperatures a high-precision agreement with heat Coulomb blockade predictions for electronic heat flow, both in the quantitative prediction of a universal suppression of exactly one electronic channel whatever $N$ and in the temperature power-law $\propto(T_\Omega^2-T^2)$.
This direct demonstration of heat Coulomb blockade in the absence of electrical Coulomb blockade constitutes the central result of the present work (see Methods for additional tests establishing the robustness of this observation).

\begin{figure*}[!tbh]
\renewcommand{\figurename}{\textbf{Figure}}
\renewcommand{\thefigure}{\textbf{\arabic{figure}}}
\centering\includegraphics[width=136mm]{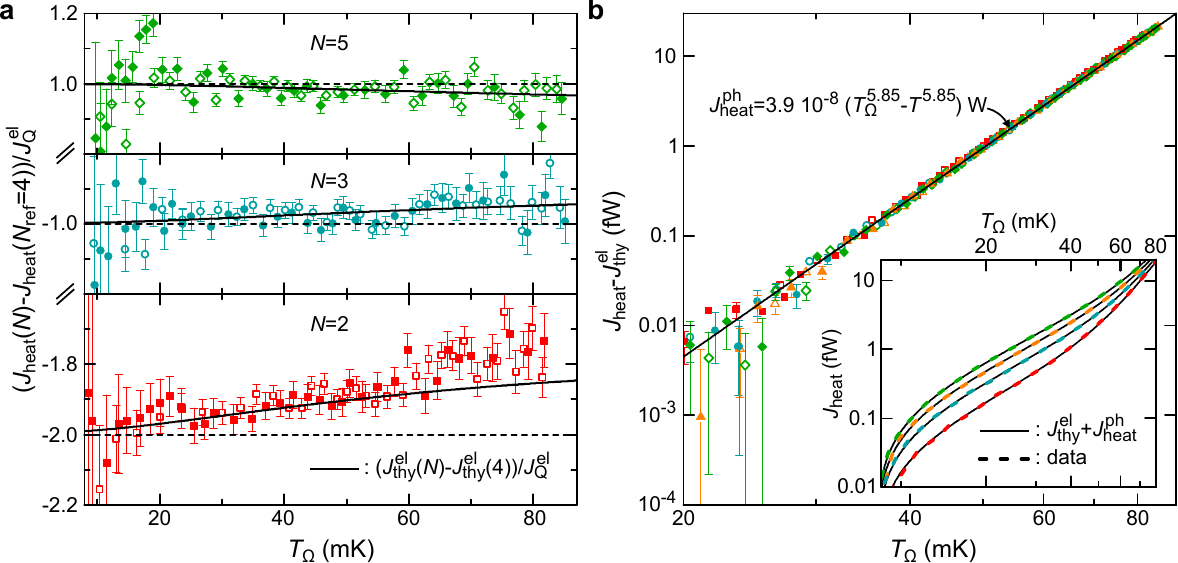}
\caption{
\footnotesize
\textbf{Heat Coulomb blockade crossover and additional mechanisms.}
\textbf{a}, Symbols (continuous lines) show the measured (predicted) heat current variation when changing $N$ from $N_\mathrm{ref}=4$ at fixed $T_\Omega$, renormalized by the quantum limit per channel $J_\mathrm{Q}^\mathrm{el}$.
The crossover toward the low-temperature heat Coulomb blockade of one ballistic channel specifically shows as a difference with respect to the nearby horizontal dashed line, whereas electron-phonon thermal transfers are canceled out.
\textbf{b}, Subtracting heat Coulomb blockade predictions, the displayed remaining part of the heat current (symbols) collapse onto a single curve for all $N\in\{2,3,4,5\}$, fitted by a $T_\Omega^{5.85}$ functional (line, $J_\mathrm{heat}^\mathrm{ph}$).
Inset, direct comparison between $J_\mathrm{thy}^\mathrm{el}+J_\mathrm{heat}^\mathrm{ph}$ (black continuous lines) and measured total heat current $J_\mathrm{heat}$ (superimposed colored dashed lines).
\normalsize
}
\label{fig3}
\end{figure*}

Theory\cite{Slobodeniuk2013} further predicts quantitatively a crossover of the electronic heat flow from $(N-1)\times J_\mathrm{Q}^\mathrm{el}$ to $N\times J_\mathrm{Q}^\mathrm{el}$, which extends mostly over one temperature decade around $NE_\mathrm{C}/\pi k_\mathrm{B}$ (Methods).
However, the relatively important and rapidly increasing thermal transfers between electrons and phonons at these higher temperatures prevent a direct observation from $J_\mathrm{heat}(T_\Omega)$.
Following Ref.~\citenum{Jezouin2013b}, we separately consider the electronic heat current by focusing on the difference at the same value of $T_\Omega$ between two numbers of connected ballistic channels ($N\in\{2,3,5\}$ and $N_\mathrm{ref}=4$).
Indeed, any mechanism that does not depend on $N$ cancels out in $J_\mathrm{heat}(N,T,T_\Omega)-J_\mathrm{heat}(N_\mathrm{ref},T,T_\Omega)$, including the electron-phonon contribution and the universal low-temperature suppression of one ballistic channel.
However, signatures of the $(N-1)\times J_\mathrm{Q}^\mathrm{el}$ to $N\times J_\mathrm{Q}^\mathrm{el}$ crossover can be observed since the temperature at which the crossover takes place increases with $N$ (Methods).
Measurements of $J_\mathrm{heat}(N\in\{2,3,5\},T,T_\Omega)-J_\mathrm{heat}(N_\mathrm{ref}=4,T,T_\Omega)$, normalized by the quantum limit of heat flow per channel $J_\mathrm{Q}^\mathrm{el}(T,T_\Omega)$, are shown as symbols versus $T_\Omega$ in Fig.~3a.
In this representation, deviations from $N-N_\mathrm{ref}$ ($N-4$, horizontal dashed lines) are specific signatures of the heat Coulomb blockade crossover.
The quantitative prediction, without any fitting parameter, of the full heat Coulomb blockade theory (continuous lines, Methods) closely matches the data.
For $N-N_\mathrm{ref}=\pm1$ ($N\in\{3,5\}$, $N_\mathrm{ref}=4$) the crossover signal is small, barely discernible at experimental accuracy, although in sign and magnitude agreement with predictions. 
For $N-N_\mathrm{ref}=-2$ ($N=2$, $N_\mathrm{ref}=4$), the larger crossover signal precisely follow the theoretical prediction up to $T_\Omega\simeq60$\,mK, while at higher $T_\Omega\gtrsim60$\,mK the scatter of the data points rapidly increases due to the overwhelming (subtracted) electron-phonon contribution $J_\mathrm{heat}^\mathrm{ph}$.
These observations further establish experimentally the full heat Coulomb blockade theory for ballistic channels at arbitrary temperatures, beyond the universal low-temperature suppression of one quantum channel.

We now investigate the additional heat transfer mechanisms at work in our device.
The main panel of Fig.~3b shows as symbols, in a $\log$-$\log$ scale versus $T_\Omega$, the measured $J_\mathrm{heat}(N)$ reduced by the heat Coulomb blockade prediction for $N$ ballistic electronic channels.
We observe that the reduced data collapse for all $N$ onto a single curve, which is closely reproduced by the functional $3.9\,10^{-8}(T_\Omega^{5.85}-T^{5.85})$\,W (continuous line).
This compares well with theoretical expectations for the electron-phonon contribution $J_\mathrm{heat}^\mathrm{ph}$ in disordered conductors, where a temperature exponent of 4 or 6 is predicted depending on the nature of disorder\cite{Sergeev2000}.
In the inset of Fig.~3b, we directly confront the measured $J_\mathrm{heat}(N\in\{2,3,4,5\},T_\Omega)$ (colored dashed lines) with the essentially indistinguishable calculations (black continuous lines) obtained by adding up the full heat Coulomb blockade prediction and the above $T_\Omega^{5.85}$ functional attributed to electron-phonon interactions.

Finally, we point out that the presently observed heat Coulomb blockade of one ballistic channel should be considered when exploiting the total outgoing heat flow from a floating node to investigate elusive exotic states\cite{Banerjee2017}.


\vspace{\baselineskip}
\small
{\noindent\textbf{Acknowledgments.}}
This work was supported by the French RENATECH network, the national French program `Investissements d'Avenir' (Labex NanoSaclay, ANR-10-LABX-0035) and the French National Research Agency (project QuTherm, ANR-16-CE30-0010-01).
We thank E.~Sukhorukov for discussions.

{\noindent\textbf{Author Contributions.}}
E.S. and F.P. performed the experiment with inputs from A.A.;
A.A., E.S. and F.P. analyzed the data;
F.D.P. fabricated the sample with inputs from A.A.;
A.C., A.O. and U.G. grew the 2DEG;
Y.J. fabricated the HEMT used for noise measurements;
F.P. led the project and wrote the manuscript with inputs from A.A., E.S. and U.G.

{\noindent\textbf{Author Information.}
Correspondence and requests for materials should be addressed to F.P. (frederic.pierre@u-psud.fr).

\normalsize

\newpage
{\Large\noindent\textbf{METHODS}}
\small

{\noindent\textbf{Sample.}} 
The sample nanostructuration is performed by standard e-beam lithography in a Ga(Al)As two-dimensional electron gas buried 105\,nm below the surface, of density $2.5\,10^{11}\,\mathrm{cm}^{-2}$ and of mobility $10^6\,\mathrm{cm}^2\mathrm{V}^{-1}\mathrm{s}^{-1}$.
The central micron-sized island is composed of a metallic multilayer of nickel (30\,nm), gold (120\,nm) and germanium (60\,nm).
Its galvanic, ohmic contact with the two-dimensional electron gas is realized by thermal annealing (440\,$^\circ$C for 50\,s).\\
The interface quality between the metallic island and the two-dimensional electron gas is fully characterized, through the individual determination of the electron reflection probability at the interface for each connected quantum Hall channel, with the self-calibrated experimental procedure detailed in Methods of Ref.~\citenum{Iftikhar2015}.
We find a reflection probability below $\lesssim0.001\%$ (the statistical uncertainty) for the three channels closest to the edge (the outer edge channel of each of the three 2DEG branches), $0.08\%$ for the inner edge channel toward electrode 3 used only for $N\in\{4,5\}$, and $0.5\%$ for the inner edge channel toward electrode 2 used only for $N=5$.\\
The typical electronic level spacing in the metallic island is estimated to be negligibly small ($\delta\approx k_\mathrm{B}\times0.2\,\mu$K), based on the electronic density of states of gold ($\nu_\mathrm{F}\approx1.14\,10^{47}$\,J$^{-1}$m$^{-3}$) and the metallic island volume ($\approx3\,\mu$m$^3$).\\
Finally, an important device parameter is the charging energy $E_\mathrm{C}\equiv e^2/2C$.
The value $E_\mathrm{C}\simeq k_\mathrm{B}\times 0.3$\,K is obtained by standard Coulomb diamond characterization, from the dc voltage height $V_\mathrm{diam}$ of the observed diamonds ($E_\mathrm{C}=eV_\mathrm{diam}/2$).
These measurements are performed in the same cooldown, at the same magnetic field (data not shown, see e.g. Fig.~1c of Ref.~\citenum{Iftikhar2016} for a similar characterization of this sample at a higher quantum Hall filling factor).\\

{\noindent\textbf{Experimental setup.}} 
The device is fixed to the mixing chamber plate of a cryofree dilution refrigerator.
Electrical measurement lines connected to the sample include several filters and thermalization stages.
Two shields at base temperature screen spurious high-frequency radiations.
Conductances are measured by standard low-frequency lock-in techniques, below 200~Hz.
Further details, including on the noise measurement setup used for the electronic thermometry, are provided in the supplementary information of Ref.~\citenum{Iftikhar2016}.\\

{\noindent\textbf{Electronic temperature.}} 
The electronic temperature in the device is extracted from on-chip quantum shot-noise measurements\cite{Iftikhar2016}.
For this purpose, we effectively short-circuit the central metallic island (equivalent circuit schematic shown bottom-left of Extended Data Fig.~1a) using the lateral continuous gate in the 2DEG branch \#2 (gate closest to the bottom in Fig.~1a).
The quantum point contact used to set $N_2$ (bottom-right split gate in Fig.~1a) is here tuned to transmit a single channel with a transmission probability $\tau\simeq0.5$.
Extended Data Fig.~1a shows as symbols the measured excess noise versus dc bias voltage $V$, and as a red continuous line the theoretical prediction at $T=7.97$\,mK for the simultaneously measured value $\tau\simeq0.515$ (variations of $\tau$ with $V$ remain below $0.003$ and are ignored).
The very low statistical uncertainty of the data shown in the main panel ($\pm5\,10^{-32}$\,A$^2/$Hz, below one tenth of the symbols size) is obtained by averaging 153 sweeps.
In practice, we fit each of the individual sweeps separately and extract from the statistical analysis of this ensemble of distinct measurements (shown as symbols in inset) the mean value of the temperature and the standard error (here $T=7.97\pm0.06$\,mK).
The same shot noise thermometry is performed both just before and just after each run of the full experiment (two runs shown in the manuscript, the first one at $T\simeq8.1$\,mK and the second at $T\simeq8.0$\,mK).\\
In addition, we consolidate the device electronic temperature $T$ with a different on-chip thermometry method based on dynamical Coulomb blockade, following  Ref.~\citenum{Iftikhar2016}.
For this purpose, the same quantum point contact (2DEG branch \#2) is set to the tunnel regime $\tau\approx0.1$ (in absence of dynamical Coulomb blockade renormalization), and the device is tuned to $N_1=0$ and $N_3=2$, effectively implementing the schematic circuit shown in the bottom-left of Extended Data Fig.~1b with a series resistance $h/2e^2$.
The temperature is obtained by fitting the conductance data displayed as symbols in the main panel of Extended Data Fig.~1b, using the known values of the series resistance and of $E_\mathrm{C}$.
We find $T=8$\,mK (corresponding calculation shown as a red line), in agreement with quantum shot noise thermometry within the larger uncertainty of dynamical Coulomb blockade thermometry, which we estimate to $\pm1$\,mK (grey area).\\

{\noindent\textbf{Gain calibration of noise amplification chain.}} 
The gain $G_\mathrm{amp}$ of the noise amplification chain is calibrated with the quantum shot noise thermometry described in section `Electronic temperature', from the linear slope of the measured shot noise at $e|V|\gg k_\mathrm{B}T$ (see  Ref.~\citenum{Iftikhar2016} for a detailed discussion).
As for the determination of $T$, we extract a different value of $G_\mathrm{amp}$ from each individual sweep of noise versus dc voltage.
From an ensemble of 409 values, $G_\mathrm{amp}$ is extracted with a statistical uncertainty of $\pm0.1\%$.
Note that the transmission probability $\tau$ enters as a factor $\tau(1-\tau)$ in the determination of $G_\mathrm{amp}$.
Here the value of $\tau$ is precisely measured simultaneously.
Although $\tau$ exhibits a weak dependence with $V$ (below $0.003$), it is sufficiently small to have a negligible impact on $G_\mathrm{amp}$ at the $\approx0.1\%$ level (note the particularly low impact in the vicinity of $\tau=0.5$) and was not taken into account.\\
The $G_\mathrm{amp}$ calibration was consolidated at an uncertainty level of $\approx1\%$, by additional quantum shot noise measurements at $\tau\simeq0.16$ (for $T\simeq8$\,mK) and also at the higher electronic temperature $T\simeq16$\,mK (for $\tau\simeq0.5$).
The comparison between the different quantum shot noise and dynamical Coulomb thermometry described in the section `Electronic temperature' further establishes the absolute calibration of $G_\mathrm{amp}$, although at a less precise level of $\approx10\%$.
Finally, we point out that the low-temperature heat Coulomb blockade reduction of heat current by precisely one $J_\mathrm{Q}^\mathrm{el}$ was also observed at the different integer quantum Hall filling factors $\nu=3$ and 4, using specific quantum shot noise calibrations for the modified gain of the noise amplification chain (with our on-chip current to voltage conversion based on the quantum Hall resistance $1/\nu G_\mathrm{Q}^\mathrm{e}$, a larger $\nu$ therefore results in a lower $G_\mathrm{amp}$). \\

{\noindent\textbf{Dissipated Joule power.}} 
The expression $P_\mathrm{J}=V^2G_\mathrm{P}/2$ is used to calculate the Joule power dissipated in the electronic fluid of the floating metallic node, due to the applied dc bias voltage $V$ (see Fig.~1a).
It corresponds to one half of the standard two-terminal total power $V^2G_\mathrm{P}$ provided by the voltage generator.
This expression can be straightforwardly derived for non-interacting electrons, within the Landauer-B\"uttiker scattering formalism.
Essentially, it amounts to sharing equally the dissipation between cold electrodes, on the one hand, and central island, on the other hand, as expected in symmetric configurations.
In the presence of interactions, one might wonder if the floating character of the central metallic node could possibly break this symmetry.
However, the dc voltage of the central island as well as the dc current across it remain unchanged compared to their free-electron values, at experimental accuracy (with the island dc voltage deduced from the measured emitted current through the quantum limited electrical conductance of the connected ballistic channels, see top-left inset of Fig.~2a) and in agreement with theory for the present case of ballistic channels.
This provides further evidence that the resulting Joule power dissipated into the central metallic island retains here its standard, non-interacting expression $P_\mathrm{J}=V^2G_\mathrm{P}/2$.
We also point out that the same non-interacting expression for $P_\mathrm{J}$ is specifically expected, on qualitative grounds (a direct quantitative derivation is yet to be done), within the theoretical heat Coulomb blockade framework for ballistic channels of Ref.~\citenum{Slobodeniuk2013} (Eugene Sukhorukov, private communication).
\\

{\noindent\textbf{Heat Coulomb blockade predictions.}} 
We provide the theoretical expression $J_\mathrm{thy}^\mathrm{el}$ derived from Ref.~\citenum{Slobodeniuk2013} for the net outgoing heat current $J_\mathrm{heat}^\mathrm{el}$ through $N$ ballistic electronic channels connecting a floating metallic node of charging energy $E_\mathrm{C}$ and at the temperature $T_\Omega$, to large electrodes at temperature $T$, for arbitrary values of $T_\Omega$ and $T$:
\begin{multline}
J_\mathrm{thy}^\mathrm{el}(N,T_\Omega,T,E_\mathrm{C})=N\frac{\pi^2k_\mathrm{B}^2}{6h}(T_\Omega^2-T^2) \\
+\frac{N^2E_\mathrm{C}^2}{\pi^2h}\left[I\left( \frac{NE_\mathrm{C}}{\pi k_\mathrm{B}T}  \right) -I\left( \frac{NE_\mathrm{C}}{\pi k_\mathrm{B}T_\Omega} \right) \right], \label{eqSLS}
\end{multline}
with the function $I$ given by
\begin{equation}
I(x)=\frac{1}{2}\left[ \ln\left(\frac{x}{2\pi}\right) - \frac{\pi}{x} -\psi\left( \frac{x}{2\pi} \right) \right],\label{eqI}
\end{equation}
where $\psi(z)$ is the digamma function.
All the displayed theoretical predictions of the full heat Coulomb blockade theory (continuous lines in Fig.~3a, inset of Fig.~3b, Extended Data Fig.~2, Extended Data Fig.~3a, inset of Extended Data Fig.~3b and Extended Data Fig.~4) were calculated using Eq.~\ref{eqSLS}, without any fitting parameter
Note that these predictions assume a hot Fermi function characterized by the temperature $T_\Omega$ for the distribution probability of the electrons in the metallic island.
This is expected since the average dwell time of the electrons in the metallic island $\tau_\mathrm{D}=h/\delta N\simeq200/N$\,$\mu$s (see e.g. Ref.~\citenum{Brower1997}) is estimated to be much larger, by about four orders of magnitude, than the typical timescale of $\sim10$\,ns for electron-electron inelastic collisions in similar metals (see e.g. Ref.~\citenum{Pierre2003} for the connected measurement of the electron coherence time in gold).\\ 

{\noindent\textbf{Asymptotic limits of predictions.}} 
The function $I$ in Eqs.~\ref{eqSLS} and \ref{eqI} has the asymptotic forms
\begin{equation}
I(x\ll1)\simeq\frac{\pi}{2x},~I(x\gg1)\simeq\frac{\pi^2}{6x^2},
\end{equation}
with a crossover centered on $x\approx1$.\\
At $T,T_\Omega\ll NE_\mathrm{C}/\pi k_\mathrm{B}$, Eq.~\ref{eqSLS} therefore reduces to
\begin{equation}
J_\mathrm{thy}^\mathrm{el}\simeq(N-1)\frac{\pi^2k_\mathrm{B}^2}{6h}(T_\Omega^2-T^2)=(N-1)\times J_\mathrm{Q}^\mathrm{el},
\end{equation}
with precisely one electronic channel effectively suppressed for heat conduction.\\
At $NE_\mathrm{C}/\pi k_\mathrm{B}\ll T,T_\Omega$, Eq.~\ref{eqSLS} reads
\begin{equation}
J_\mathrm{thy}^\mathrm{el}\simeq N\frac{\pi^2k_\mathrm{B}^2}{6h}(T_\Omega^2-T^2)-N\frac{E_\mathrm{C}k_\mathrm{B}}{2h}(T_\Omega-T),
\end{equation}
which corresponds to a net reduction of the heat conductance ($|T-T_\Omega|\rightarrow0$) per ballistic electronic channel by the fixed amount $\Delta G_\mathrm{heat}^\mathrm{el}=E_\mathrm{C}k_\mathrm{B}/2h$ (always small with respect to $G_\mathrm{Q}^\mathrm{h}$ in the considered high-temperature limit).\\
At $T\ll NE_\mathrm{C}/\pi k_\mathrm{B}\ll T_\Omega$, Eq.~\ref{eqSLS} becomes
\begin{equation}
J_\mathrm{thy}^\mathrm{el}\simeq \frac{\pi^2k_\mathrm{B}^2}{6h}(NT_\Omega^2-(N-1)T^2)-N\frac{E_\mathrm{C}k_\mathrm{B}}{2h}T_\Omega,
\end{equation}
where the relative reduction due to heat Coulomb blockade progressively vanishes as $T_\Omega$ increases.\\

{\noindent\textbf{Control experiment at \textit{T}$\mathbf{\simeq}$16\,mK.}} 
We here demonstrate the robustness of our result with respect to base temperature $T$.
The experiment is performed at a temperature twice as large as before, $T\simeq15.9\pm0.1$\,mK, for the setting $N=2$.
As seen from the $J_\mathrm{heat}$ vs $T_\Omega^2-T^2$ data shown as symbols in Extended Data Fig.~3a, this is still sufficiently low to directly and quantitatively establish the heat Coulomb blockade suppression of one ballistic channel. 
Note that this robustness also further validates the specific tests performed to rule out possible experimental artifacts (additional power injection by the measurement lines, voltage-dependent noise offset, calibration and thermometry issues).\\
We also find that exactly the same functional $3.9\,10^{-8}(T_\Omega^{5.85}-T^{5.85})$, attributed to electron-phonon (black continuous line in main panel of Extended Data Fig.~3b), matches the present $T\simeq16$\,mK data reduced by the predicted electronic heat flow (symbols in main panel of Extended Data Fig.~3b).
Note that the observed independence of the electron-phonon coupling with temperature is a commonly used criterion to show that cold electrons and phonons are at the same temperature $T$ (see e.g. Ref.~\citenum{Rajauria2007}).\\

{\noindent\textbf{Supplementary experimental tests.}} 
We also performed the following tests:\\
(\textit{i}) At $N=0$, the measured noise is found to be independent of dc bias (applied either within the same 2DEG branch or in a disconnected branch, with the current flowing toward cold grounds not shown in Fig.~1), at experimental accuracy.
In the language of Ref.~\citenum{Banerjee2017}, the `source noise' is negligible.\\
(\textit{ii}) At $V=0$, the absolute measured noise does not change when tuning $N$ to different values or by connecting the amplification noise chain to an edge channel emitted from a cold ground.
This directly shows that the temperature $T$ is homogeneous (the same $T$ for all large electrodes and the metallic island, whatever $N$), and also that the power injected into the central island is negligible at $V=0$ (in the presence of e.g. significant heating from the measurement lines, the central island temperature and consequently the noise measured would depend on $N_{1,2,3}$; note that such an heating takes place in one of the test configurations discussed in (\textit{iii})).\\
(\textit{iii}) We checked that the observed heat Coulomb blockade is independent of the specific channel realization, by comparing three different device configurations all corresponding to $N=2$ (see Extended Data Fig.~4): 
$(N_1=1,N_2=1,N_3=0)$ (main manuscript, red squares in Extended Data Fig.~4), $(N_1=0,N_2=1,N_3=1)$ (violet squares in Extended Data Fig.~4) and $(N_1=1,N_2=0,N_3=1)$ (black squares in Extended Data Fig.~4).
The first two configurations give exactly the same result at experimental accuracy.
The third configuration has the additional complication that, in order to perform the noise thermometry of the central island, we had to make use of an otherwise disconnected measurement line within the 2DEG branch 3, through which a non-negligible power was dissipated into the island even at $V=0$ (as seen with measurements as discussed in (\textit{ii}), and also with other tests including quantum shot noise and dynamical Coulomb blockade measurements).
Nonetheless, heat Coulomb blockade predictions (and a full agreement with the first two device configurations) are verified when taking into account the separately calibrated temperature offset $T_\Omega(V=0)-T\simeq5$\,mK (obtained through the approach of test (\textit{ii})) and the additional dissipated power of $0.06$\,fW at $V=0$.\\
(\textit{iv}) Finally, we verified the robustness of our results with respect to the integer quantum Hall filling factor $\nu$ at which the experiment is performed.
The heat Coulomb blockade suppression of one ballistic channel was quantitatively observed experimentally not only at $\nu=2$ (main manuscript), but also at $\nu=3$ and $\nu=4$.\\

{\noindent\textbf{Comparison with previous experiments.}} 
In the previous works Refs~\citenum{Jezouin2013b,Banerjee2017}, besides the robust $\Delta J_\mathrm{heat}$ approach to extract the quantum limit of electronic heat flow $J_\mathrm{Q}^\mathrm{el}$, it was also attempted to determine the full electronic heat flow despite the non-negligible electron-phonon contribution at $T\gtrsim25$\,mK.
For this purpose, a $T_\Omega^5-T^5$ power-law was assumed to model the transfer of energy from electrons at $T_\Omega$ toward phonons at $T$.
With this model, no heat Coulomb blockade was detected\cite{Jezouin2013b,Banerjee2017}.
However, we here point out the sensitive influence of the electron-phonon power law exponent in the analysis of these previous experiments.
Together with the higher $T$, it essentially impeded the observation of the heat Coulomb blockade.
We illustrate this with the most accurate data of Ref.~\citenum{Jezouin2013b}, obtained at integer quantum Hall filling factor $\nu=3$.
As pointed out in the supplementary materials of Ref.~\citenum{Jezouin2013b}, fitting the data for reference channel number $N_\mathrm{ref}=4$ of this previous experiment with the electron-phonon temperature exponent as a free parameter (instead of assuming an exponent of 5) leads to a reduced overall electronic heat flow of $\approx3.5\times J_\mathrm{Q}^\mathrm{el}$, a factor in between $N_\mathrm{ref}$ and $N_\mathrm{ref}-1$, predicted respectively in the absence and in the presence of a heat Coulomb blockade.
Reanalyzing these previous data with the electron-phonon power-law left as a free parameter, we find that they are indeed compatible at experimental accuracy with the present observation of the heat Coulomb blockade of one ballistic channel.\\

\normalsize

\newpage

\begin{figure*}[t]
\renewcommand{\figurename}{\textbf{Extended Data Figure}}
\renewcommand{\thefigure}{\textbf{1}}
\centering\includegraphics [width=1.8\columnwidth]{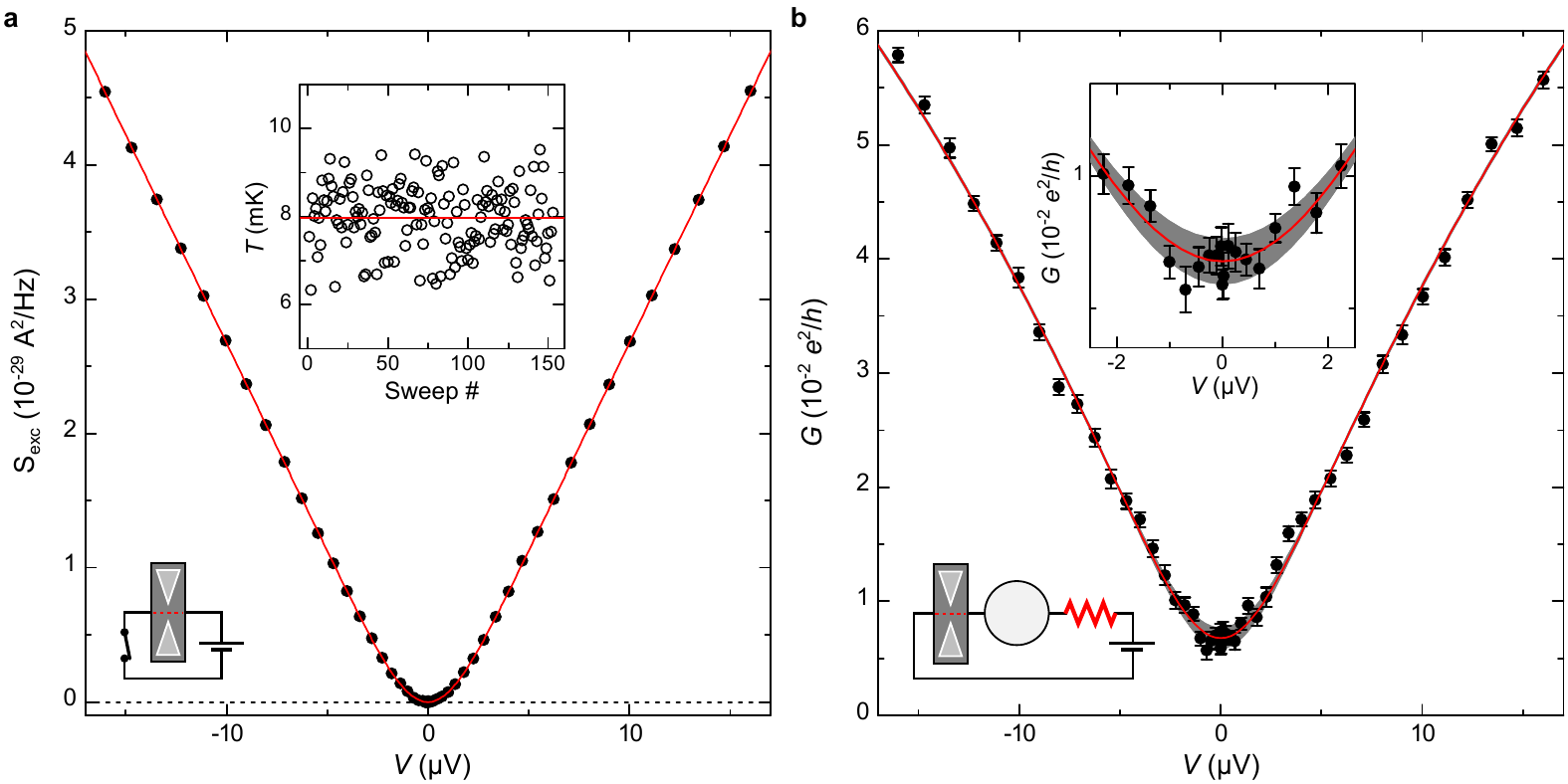}
\caption{
\footnotesize
\textbf{Electronic temperature.}
\textbf{a}, Quantum shot noise thermometry.
Symbols in the main panel represent the measured excess spectral density of the current fluctuations across a quantum point contact set to transmit a single electronic channel with a probability $\tau\simeq0.516$, and biased with the dc voltage $V$ (see configuration schematic).
The statistical uncertainty of $\pm5\,10^{-32}$\,A$^2$/Hz on $S_\mathrm{exc}$ is below one tenth of the symbols size.
The red continuous line is the calculated excess current fluctuations for $T=7.97$\,mK and $\tau=0.516$.
Inset: the different electronic temperatures $T$ shown as symbols are each obtained by fitting a different (successive) voltage bias sweep of the quantum shot noise (symbols in the main panel represent the average of these sweeps).
From the statistical analysis of these 153 values, we find $T\simeq7.97\pm0.06$\,mK (horizontal red line).
\textbf{b}, Dynamical Coulomb blockade thermometry.
The electronic temperature is here obtained by fitting the device conductance $G$ (symbols) versus voltage bias with the dynamical Coulomb blockade theory in the presence of a known series resistance $R=h/2e^2$ (see configuration schematic).
We find $T\simeq8$\,mK from the fit shown as a continuous line.
The estimated uncertainty of $\pm1$\,mK is displayed as a grey background best visible in the inset showing only low voltages.
}
\end{figure*}

\begin{figure*}[b]
\renewcommand{\figurename}{\textbf{Extended Data Figure}}
\renewcommand{\thefigure}{\textbf{2}}
\centering\includegraphics[width=2\columnwidth]{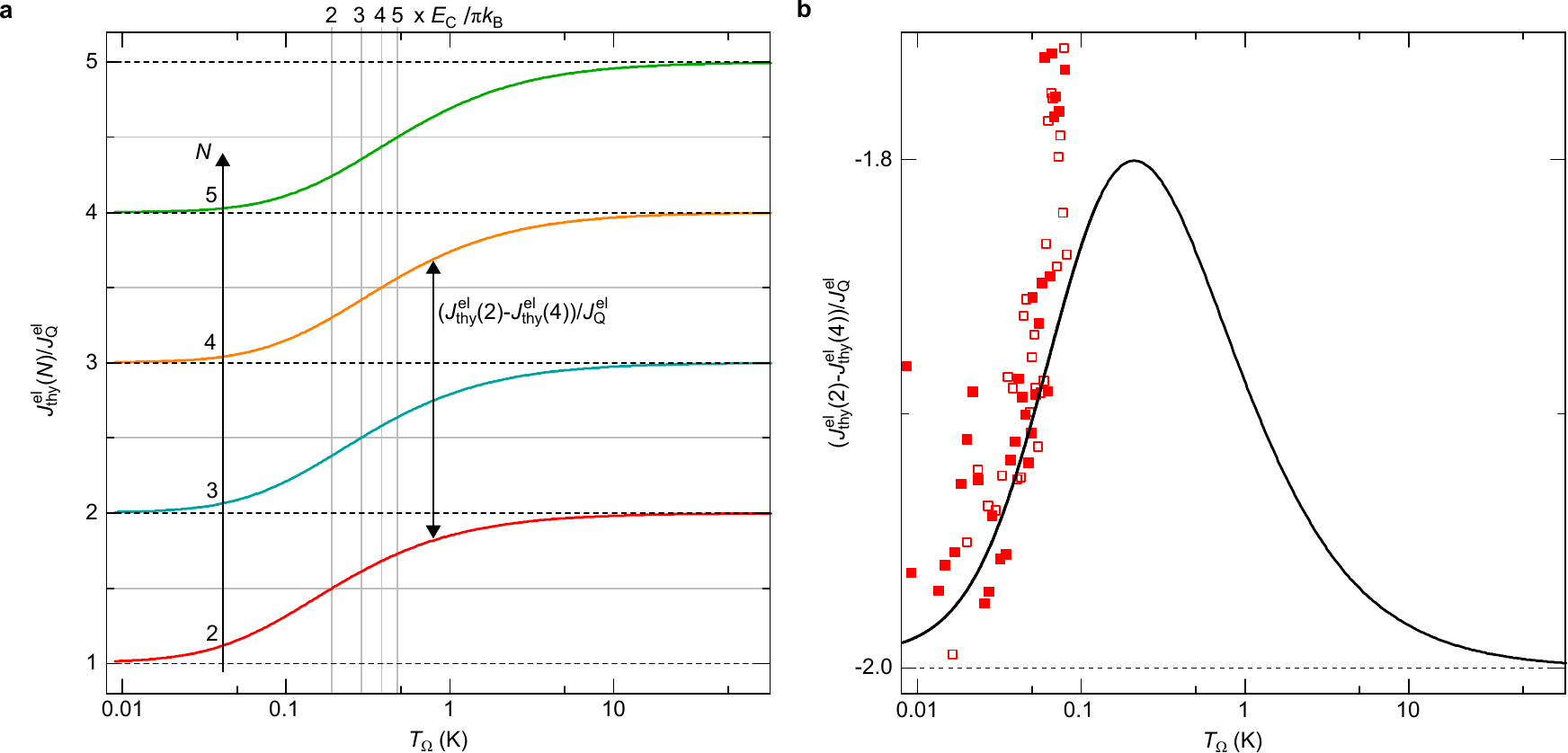}
\caption{
\footnotesize
\textbf{Heat Coulomb blockade versus temperature.}
\textbf{a}, Crossover from $N\times J_\mathrm{Q}^\mathrm{el}$ at $T\gg N E_\mathrm{C}/\pi k_\mathrm{B}$ to $(N-1)\times J_\mathrm{Q}^\mathrm{el}$ at $T\ll N E_\mathrm{C}/\pi k_\mathrm{B}$.
Colored continuous lines display the theoretical prediction of Eq.~\ref{eqSLS} ($T=8.07\,$mK, $E_\mathrm{C}=0.3\times k_\mathrm{B}\,$K) normalized by the quantum limit of heat flow per channel $J_\mathrm{Q}^\mathrm{el}$, for different numbers of ballistic channels $N\in\{2,3,4,5\}$ and versus the metallic node temperature $T_\Omega$ in $\log$ scale.
\textbf{b}, The normalized difference between heat flows at $N=2$ and $N=4$ (double arrow in \textbf{a}, previously shown in the bottom of Fig.~3a on the experimentally explored temperature range) is displayed on the full temperature range where the crossover takes place.
The black continuous line is the prediction of Eq.~\ref{eqSLS}.
The same data previously shown in the bottom part of Fig.~3a are displayed as symbols.
}
\end{figure*}

\begin{figure*}[t]
\renewcommand{\figurename}{\textbf{Extended Data Figure}}
\renewcommand{\thefigure}{\textbf{3}}
\centering\includegraphics[width=2\columnwidth]{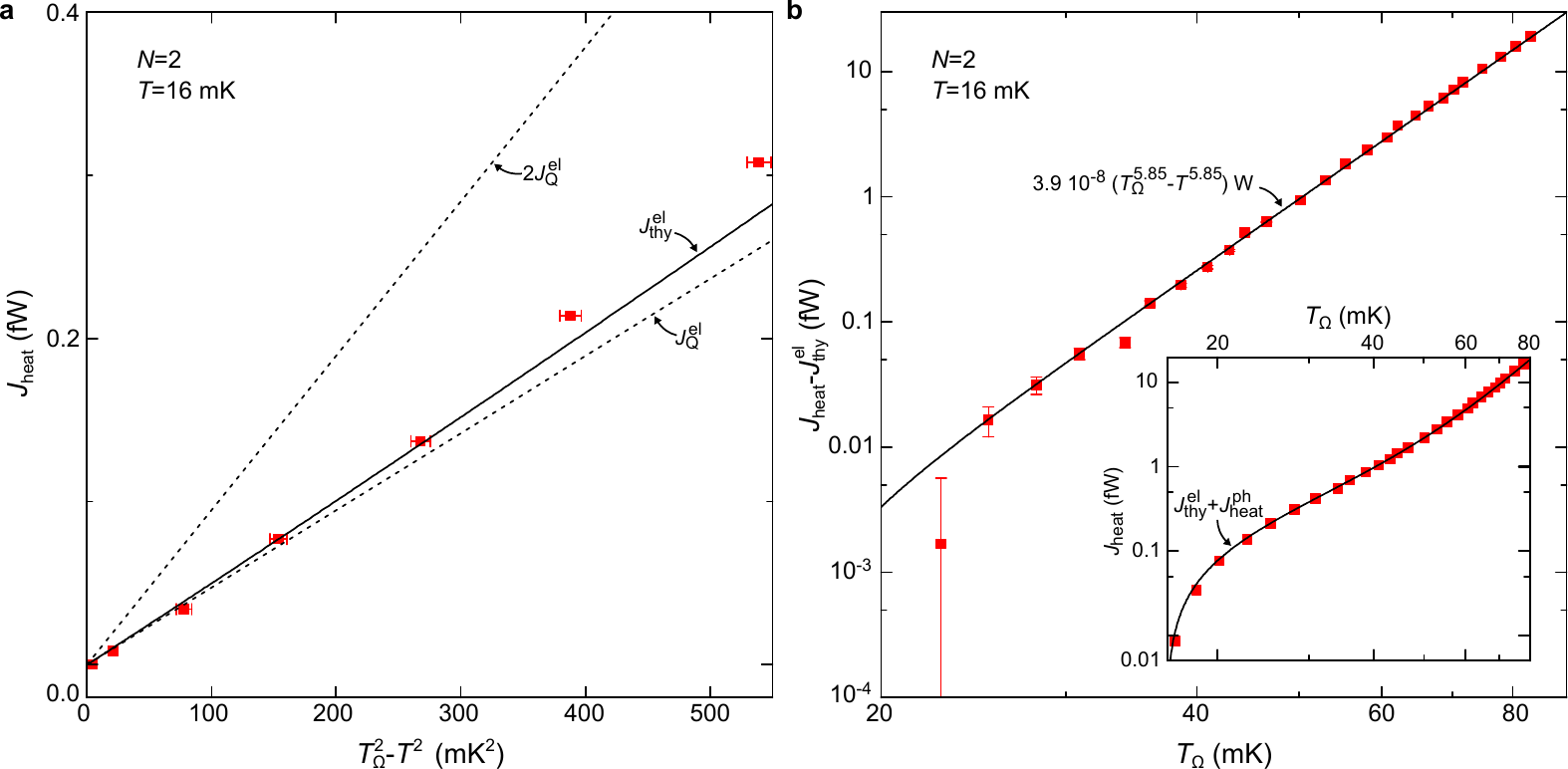}
\caption{
\footnotesize
\textbf{Control experiment at \textit{T}$\mathbf{\simeq16}$\,mK.}
\textbf{a}, Heat Coulomb blockade of one ballistic channel.
Measurements at $N=2$ of the overall heat flow ($J_\mathrm{heat}=P_\mathrm{J}$) versus $T_\Omega^2-T^2$ are displayed as symbols, for low temperatures where the electron-phonon contribution remains relatively small ($T_\Omega<30$\,mK).
The bottom dashed line, close to low-temperature data points, corresponds to the low-temperature asymptotic suppression of precisely one $J_\mathrm{Q}^\mathrm{el}$.
The quantitative heat Coulomb blockade prediction of Eq.~\ref{eqSLS} for the electronic thermal transport is shown as a continuous line.
The top dashed line corresponds to the prediction for the electronic thermal current across two ballistic channels in the absence of heat Coulomb blockade, $2J_\mathrm{Q}^\mathrm{el}$. 
\textbf{b}, Additional mechanisms.
As in Fig.~3b, we display as symbols the measured heat current reduced by the quantitative heat Coulomb blockade prediction ($N=2$, $T=15.9$\,mK).
We observe an exact match with the same $T_\Omega^{5.85}$ functional (line) obtained at the base temperature of 8\,mK.
Inset, the sum of heat Coulomb blockade predictions and $T_\Omega^{5.85}$ functional (black continuous line) is directly compared to measured $J_\mathrm{heat}(T_\Omega)$ (symbols).
}
\end{figure*}

\begin{figure*}[!b]
\renewcommand{\figurename}{\textbf{Extended Data Figure}}
\renewcommand{\thefigure}{\textbf{4}}
\centering\includegraphics [width=1\columnwidth]{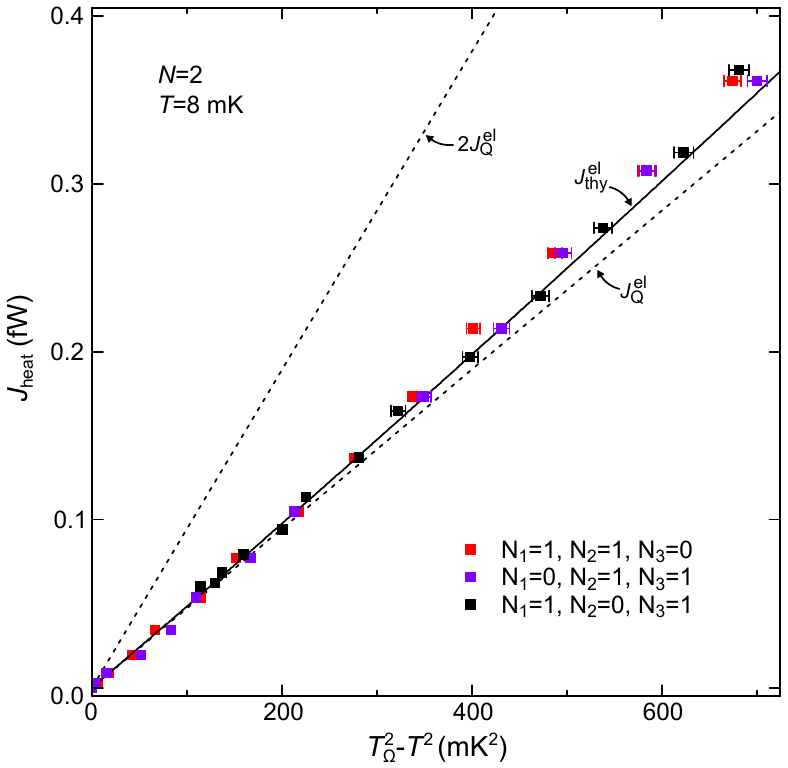}
\caption{
\footnotesize
\textbf{Comparison of three device configurations implementing \textit{N}=2.}
The three sets of symbols, each displayed with a different color, correspond to measurements of $J_\mathrm{heat}$ vs $T_\Omega^2-T^2$ with the device tuned in different configurations $\left[N_1,N_2,N_3\right]$ (as detailed in the figure), all associated with the same total number of ballistic channels connected to the node ($N=2$).
The bottom and top dashed lines correspond, respectively, to $J_\mathrm{Q}^\mathrm{el}$ and $2\times J_\mathrm{Q}^\mathrm{el}$.
The continuous line is the quantitative heat Coulomb blockade prediction for electronic thermal transport given by Eq.~\ref{eqSLS}. 
}
\end{figure*}


\begin{thebibliography}{33}
\expandafter\ifx\csname natexlab\endcsname\relax\def\natexlab#1{#1}\fi
\expandafter\ifx\csname url\endcsname\relax
  \def\url#1{\texttt{#1}}\fi
\expandafter\ifx\csname urlprefix\endcsname\relax\def\urlprefix{URL }\fi

\bibitem[{Pekola(2015)}]{Pekola2015}
Pekola, J.
\newblock Towards quantum thermodynamics in electronic circuits.
\newblock \emph{Nat. Phys.} \textbf{11}, 118--123 (2015).

\bibitem[{Vinjanampathy \& Anders(2016)}]{Vinjanampathy2016}
Vinjanampathy, S. \& Anders, J.
\newblock Quantum thermodynamics.
\newblock \emph{Contemp. Phys} \textbf{57}, 545--579 (2016).

\bibitem[{Schwab \emph{et~al.}(2000)Schwab, Henriksen, Worlock \&
  Roukes}]{Schwab2000}
Schwab, K., Henriksen, E., Worlock, J. \& Roukes, M.
\newblock {Measurement of the quantum of thermal conductance}.
\newblock \emph{Nature} \textbf{404}, 974--977 (2000).

\bibitem[{Meschke \emph{et~al.}(2006)Meschke, Guichard \& Pekola}]{Meschke2006}
Meschke, M., Guichard, W. \& Pekola, J.~P.
\newblock {Single-mode heat conduction by photons}.
\newblock \emph{Nature} \textbf{444}, 187--190 (2006).

\bibitem[{Jezouin \emph{et~al.}(2013{\natexlab{a}})}]{Jezouin2013b}
Jezouin, S. \emph{et~al.}
\newblock Quantum Limit of Heat Flow Across a Single Electronic Channel.
\newblock \emph{Science} \textbf{342}, 601--604 (2013{\natexlab{a}}).

\bibitem[{{Banerjee} \emph{et~al.}(2017)}]{Banerjee2017}
{Banerjee}, M. \emph{et~al.}
\newblock Observed Quantization of Anyonic Heat Flow.
\newblock \emph{Nature} \textbf{545}, 75--79 (2017).

\bibitem[{Cui \emph{et~al.}(2017)}]{Cui2017}
Cui, L. \emph{et~al.}
\newblock Quantized thermal transport in single-atom junctions.
\newblock \emph{Science} \textbf{355}, 1192--1195 (2017).

\bibitem[{Giazotto \& Mart\'{\i}nez-P\'{e}rez(2012)}]{Giazotto2012}
Giazotto, F. \& Mart\'{\i}nez-P\'{e}rez, M.~J.
\newblock {The Josephson heat interferometer}.
\newblock \emph{Nature} \textbf{492}, 401--405 (2012).

\bibitem[{Slobodeniuk \emph{et~al.}(2013)Slobodeniuk, Levkivskyi \&
  Sukhorukov}]{Slobodeniuk2013}
Slobodeniuk, A., Levkivskyi, I. \& Sukhorukov, E.
\newblock Equilibration of quantum Hall edge states by an Ohmic contact.
\newblock \emph{Phys. Rev. B} \textbf{88}, 165307 (2013).

\bibitem[{Landauer(1975)}]{Landauer1975}
Landauer, R.
\newblock Residual resistivity dipoles.
\newblock \emph{Z. Phys. B} \textbf{21}, 247--254 (1975).

\bibitem[{B\"uttiker(1986)}]{Buttiker1986}
B\"uttiker, M.
\newblock Four-Terminal Phase-Coherent Conductance.
\newblock \emph{Phys. Rev. Lett.} \textbf{57}, 1761--1764 (1986).

\bibitem[{Kulik \& Shekhter(1975)}]{Kulik1975}
Kulik, I. \& Shekhter, R.
\newblock Kinetic phenomena and charge discreteness effects in granulated
  media.
\newblock \emph{Sov. Phys. JETP} \textbf{41}, 308--316 (1975).

\bibitem[{Averin \& Likharev(1986)}]{Averin1986}
Averin, D. \& Likharev, K.
\newblock Coulomb blockade of single-electron tunneling, and coherent
  oscillations in small tunnel junctions.
\newblock \emph{J. Low Temp. Phys.} \textbf{62}, 345--373 (1986).

\bibitem[{Nazarov(1989)}]{Nazarov1989}
Nazarov, Y.
\newblock Anomalous current-voltage characteristics of tunnel junctions.
\newblock \emph{Sov. Phys. JETP} \textbf{68}, 561--566 (1989).

\bibitem[{Grabert \& Devoret(1992)}]{Ingold1992}
Grabert, H. \& Devoret, M.~H. (eds.).
\newblock \emph{Single charge tunneling} (1992), plenum, new york edn.

\bibitem[{Matveev(1991)}]{Matveev1991}
Matveev, K.~A.
\newblock Quantum fluctuations of the charge of a metal particle under the
  Coulomb blockade conditions.
\newblock \emph{Sov. Phys. JETP} \textbf{72}, 892--899 (1991).

\bibitem[{Iftikhar \emph{et~al.}(2015)}]{Iftikhar2015}
Iftikhar, Z. \emph{et~al.}
\newblock Two-channel Kondo effect and renormalization flow with macroscopic
  quantum charge states.
\newblock \emph{Nature} \textbf{526}, 233--236 (2015).

\bibitem[{Flensberg(1993)}]{Flensberg1993}
Flensberg, K.
\newblock Capacitance and conductance of mesoscopic systems connected by
  quantum point contacts.
\newblock \emph{Phys. Rev. B} \textbf{48}, 11156--11166 (1993).

\bibitem[{Yeyati \emph{et~al.}(2001)Yeyati, Martin-Rodero, Esteve \&
  Urbina}]{Yeyati2001}
Yeyati, A.~L., Martin-Rodero, A., Esteve, D. \& Urbina, C.
\newblock Direct Link between Coulomb Blockade and Shot Noise in a
  Quantum-Coherent Structure.
\newblock \emph{Phys. Rev. Lett.} \textbf{87}, 046802--046805 (2001).

\bibitem[{Kindermann \& Nazarov(2003)}]{Kindermann2003}
Kindermann, M. \& Nazarov, Y.~V.
\newblock Interaction Effects on Counting Statistics and the Transmission
  Distribution.
\newblock \emph{Phys. Rev. Lett.} \textbf{91}, 136802--136805 (2003).

\bibitem[{Altimiras \emph{et~al.}(2007)Altimiras, Gennser, Cavanna, Mailly \&
  Pierre}]{Altimiras2007}
Altimiras, C., Gennser, U., Cavanna, A., Mailly, D. \& Pierre, F.
\newblock Experimental Test of the Dynamical Coulomb Blockade Theory for Short
  Coherent Conductors.
\newblock \emph{Phys. Rev. Lett.} \textbf{99}, 256805 (2007).

\bibitem[{Parmentier \emph{et~al.}(2011)}]{Parmentier2011}
Parmentier, F.~D. \emph{et~al.}
\newblock Strong back-action of a linear circuit on a single electronic quantum
  channel.
\newblock \emph{Nat. Phys.} \textbf{7}, 935--938 (2011).

\bibitem[{Jezouin \emph{et~al.}(2013{\natexlab{b}})}]{Jezouin2013}
Jezouin, S. \emph{et~al.}
\newblock Tomonaga-Luttinger physics in electronic quantum circuits.
\newblock \emph{Nat. Comm.} \textbf{4}, 1802 (2013{\natexlab{b}}).

\bibitem[{Altimiras \emph{et~al.}(2012)}]{Altimiras2012}
Altimiras, C. \emph{et~al.}
\newblock Chargeless Heat Transport in the Fractional Quantum Hall Regime.
\newblock \emph{Phys. Rev. Lett.} \textbf{109}, 026803 (2012).

\bibitem[{Dutta \emph{et~al.}(2017)}]{Dutta2017}
Dutta, B. \emph{et~al.}
\newblock Thermal Conductance of a Single-Electron Transistor.
\newblock \emph{Phys. Rev. Lett.} \textbf{119}, 077701 (2017).

\bibitem[{Blanter \& B\"uttiker(2000)}]{Blanter2000}
Blanter, Y.~M. \& B\"uttiker, M.
\newblock Shot Noise in Mesoscopic Conductors.
\newblock \emph{Phys. Rep.} \textbf{336}, 1--166 (2000).

\bibitem[{Liang \emph{et~al.}(2012)Liang, Dong, Gennser, Cavanna \&
  Jin}]{Liang2012}
Liang, Y., Dong, Q., Gennser, U., Cavanna, A. \& Jin, Y.
\newblock Input Noise Voltage Below 1\,nV/Hz$^{1/2}$ at 1\,kHz in the HEMTs at
  4.2\,K.
\newblock \emph{J. Low Temp. Phys.} \textbf{167}, 632--637 (2012).

\bibitem[{Iftikhar \emph{et~al.}(2016)}]{Iftikhar2016}
Iftikhar, Z. \emph{et~al.}
\newblock Primary thermometry triad at 6 mK in mesoscopic circuits.
\newblock \emph{Nat. Commun.} \textbf{7}, 12908 (2016).

\bibitem[{Blanter \& Sukhorukov(2000)}]{Blanter2000chaotic}
Blanter, Y.~M. \& Sukhorukov, E.~V.
\newblock Semiclassical Theory of Conductance and Noise in Open Chaotic
  Cavities.
\newblock \emph{Phys. Rev. Lett.} \textbf{84}, 1280--1283 (2000).

\bibitem[{Sergeev \& Mitin(2000)}]{Sergeev2000}
Sergeev, A. \& Mitin, V.
\newblock {Electron-phonon interaction in disordered conductors: Static and
  vibrating scattering potentials}.
\newblock \emph{Phys. Rev. B} \textbf{61}, 6041--6047 (2000).

\bibitem[{Brouwer \& B\"uttiker(1997)}]{Brower1997}
Brouwer, P.~W. \& B\"uttiker, M.
\newblock Charge-relaxation and dwell time in the fluctuating admittance of a
  chaotic cavity.
\newblock \emph{Europhys. Lett.} \textbf{37}, 441--446 (1997).

\bibitem[{Pierre \emph{et~al.}(2003)}]{Pierre2003}
Pierre, F. \emph{et~al.}
\newblock Dephasing of electrons in mesoscopic metal wires.
\newblock \emph{Phys. Rev. B} \textbf{68}, 085413 (2003).

\bibitem[{Rajauria \emph{et~al.}(2007)}]{Rajauria2007}
Rajauria, S. \emph{et~al.}
\newblock Electron and Phonon Cooling in a Superconductor-Normal
  Metal-Superconductor Tunnel Junction.
\newblock \emph{Phys. Rev. Lett.} \textbf{99}, 047004 (2007).

\end{thebibliography}
\end{document}